\documentclass{article}
\usepackage{dcase2021,amsmath,graphicx,url,times,booktabs, tabularx}
\usepackage{amsfonts}

\usepackage{tikz}
\usepackage[most]{tcolorbox}
\usepackage{hyperref}
\usepackage{subcaption}
\usepackage{makecell}
\newcommand{\norm}[1]{\left\lVert#1\right\rVert}

\title{USING UMAP TO INSPECT AUDIO DATA FOR UNSUPERVISED ANOMALY DETECTION UNDER DOMAIN-SHIFT CONDITIONS}

\name{Andres Fernandez, Mark D. Plumbley}
\address{Centre for Vision, Speech and Signal Processing (CVSSP)\\
  University of Surrey, UK\\
      \{andres.fernandez, m.plumbley\}@surrey.ac.uk}

\usepackage[nolist,nohyperlinks, printonlyused]{acronym}
\acrodef{UMAP}[UMAP]{Uniform Manifold Approximation and Projection}
\acrodefplural{UMAP}[UMAPs]{Uniform Manifold Approximations and Projections}
\acrodef{DCASE}[DCASE]{Detection and Classification of Acoustic Scenes and Events}
\acrodef{DL}[DL]{Deep Learning}
\acrodef{RepL}[RepL]{Representation Learning}
\acrodef{FSTs}[FSTs]{fixed signal transformations}
\acrodef{UAD}[UAD]{Unsupervised Anomaly Detection}
\acrodef{UAD-S}[UAD-S]{Unsupervised Anomaly Detection under Domain-Shift Conditions}
\acrodef{AUC}[AUC]{Area Under ROC Curve}
\acrodef{pAUC}[pAUC]{Area Under Partial ROC Curve}
\acrodef{AE}[AE]{autoencoder}
\acrodef{IAE}[IAE]{Interpolation Autoencoder}
\acrodef{MSE}[MSE]{Mean Square Error}
\acrodef{BCE}[BCE]{Binary Cross-Entropy}
\acrodef{KDE}[KDE]{Kernel Density Estimation}
\acrodef{LOF}[LOF]{Local Outlier Factor}
\acrodef{KNN}[KNN]{K-Nearest Neighbors}
\acrodef{PCA}[PCA]{Principal Component Analysis}
\acrodef{LDA}[LDA]{Linear Discriminant Analysis}
\acrodef{GDA}[GDA]{Generalized Discriminant Analysis}
\acrodef{SVM}[SVM]{Support Vector Machine}
\acrodef{VAE}[VAE]{Variational Autoencoder}
\acrodef{CVAE}[CVAE]{Conditional Variational Autoencoder}
\acrodef{EAE}[EAE]{Extrapolation Autoencoder}
\acrodef{RAE}[RAE]{Reconstruction Autoencoder}
\acrodef{L3}[L3]{Look, Listen and Learn}
\acrodef{PDF}[PDF]{Probability Density Function}
\acrodef{CDF}[CDF]{Cumulative Density Function}
\acrodef{ISA}[ISA]{Industrial Sound Analysis}
\acrodef{PdM}[PdM]{Predictive Maintenance}
\acrodef{EPSRC}[EPSRC]{Engineering and Physical Sciences Research Council}
\acrodef{SEP}[SEP]{Separability}
\acrodef{DSUP}[DSUP]{Discriminative Support}
\acrodef{STFT}[STFT]{Short-Term Fourier Transform}

\begin{document}

\ninept
\maketitle

\begin{sloppy}

  \begin{abstract}
    The goal of \ac{UAD} is to detect anomalous signals under the condition that only non-anomalous ({\it normal}) data is available beforehand. In UAD under Domain-Shift Conditions (UAD-S), data is further exposed to contextual changes that are usually unknown beforehand. Motivated by the difficulties encountered in the \acs{UAD-S} task presented at the 2021 edition of the \ac{DCASE} challenge\footnote{DCASE website: \url{http://dcase.community}}, we visually inspect \acp{UMAP} for log-STFT, log-mel and pretrained \ac{L3} representations of the \ac{DCASE} \acs{UAD-S} dataset. In our exploratory investigation, we look for two qualities, {\it \ac{SEP}} and {\it \ac{DSUP}}, and formulate several hypotheses that could facilitate diagnosis and developement of further representation and detection approaches. Particularly, we hypothesize that input length and pretraining may regulate a relevant tradeoff between \ac{SEP} and \ac{DSUP}. Our code as well as the resulting \acp{UMAP} and plots are publicly available\footnote{Online Resources:\\ Code: \url{https://github.com/andres-fr/dcase2021_umaps}\\Webpage: \url{https://ai4s.surrey.ac.uk/2021/dcase_uads}}.
\end{abstract}
 

\acresetall  % reset acronyms

\section{Introduction}

The goal of \ac{UAD} is to detect anomalous instances under the condition that only non-anomalous (i.e.\ {\it normal}) instances are available beforehand. This has relevance in monitoring applications where anomalous data is hard to collect whereas normal data is abundant. \ac{UAD-S} presents an extra challenge: both normal and anomalous data can be exposed to {\it domain shifts}, i.e.\ changes in the environment that cause an impact in the data and are usually unknown beforehand. This can result in false negatives, if the detector is not sensitive enough, or false positives, if the detector does not tolerate or adapt to domain shifts.\\
In the audio domain, \ac{UAD} has attracted attention as a promising \ac{PdM}\cite{pdm} solution for \ac{ISA}\cite{isa}: Sound monitoring is non-invasive, is robust to occlusions, can be carried out during production, and anomalous sounds can signal issues long before critical faults occur. \ac{UAD-S} is a natural extension to \ac{UAD}, since even in controlled environments, new non-anomalous sources of sound can arise (e.g.\ due to maintenance work or upgrades). Given that it is difficult or undesirable to completely isolate the analyzed sound source from its environment, \ac{PdM}-\ac{ISA} solutions must be able to detect slight deviations (including short-duration events like clicks) while embracing stronger environmental changes.\\
The 2020 and 2021 editions of \ac{DCASE} have incorporated \ac{UAD} (2020, task 2) and \ac{UAD-S} (2021, task 2) challenges. In both editions, a broad variety of approaches has been explored, but the results achieved in 2021 were significantly lower than in 2020 in terms of numeric performance. This may indicate a higher complexity of the 2021 task, independently of the choice of model and training scheme. In order to gain further insights, we propose to inspect the data distribution itself. Specifically, our proposed contributions are:

\begin{itemize}
\item We showcase a method for of \ac{UAD-S} data exploration via visual inspection of \acp{UMAP} and assessment of 2 beneficial qualities: {\bf \ac{SEP}} and {\bf \ac{DSUP}}.
\item We apply the proposed analysis procedure to the \ac{DCASE} 2021 dataset, revealing insights on its macro- and microstructure.
\item Based on the analysis and literature, we formulate a series of verifiable hypotheses that we believe can facilitate diagnosis and developement of further approaches.
\end{itemize}
Section \ref{sec:dcase} reviews \ac{UAD} in \ac{DCASE}. Section \ref{sec:methodology} describes our methodology. Section \ref{sec:pipeline} describes our experiments. Section \ref{sec:discussion} presents and discusses some results. Section \ref{sec:conclusion} concludes and proposes future work.

\section{UAD in DCASE}
\label{sec:dcase}

The \ac{DCASE} 2020 dataset was a result of combining two recently curated datasets (ToyAdmos\cite{toyadmos} and MIMII\cite{mimii}), each featuring 10-second audio segments from different well-functioning {\it devices} (toy car, valve, fan, etc). Each segment was mixed with different background sounds to simulate real environments. The devices were then intentionally damaged/disrupted to provide anomalous data, which was only available for validation. The proposed models had to provide a real-valued anomaly score for each validation audio segment, and their performance was evaluated by ranking the \ac{AUC} and \ac{pAUC} obtained across different devices\cite{2020t2_summary}.\\
The 10 best performing submissions in 2020 applied \ac{DL}, treating the developement dataset as training data. Some used different forms of data augmentation and additions from external datasets such as AudioSet\cite{audioset} and Fraunhofer's IDMT-ISA-EE dataset\cite{isa}. For inference, most submissions directly applied the trained \ac{DL} models, often via multi-task ensembles. The most popular alternative was to apply \ac{KNN} to learned embeddings of the training set\cite{arcface_zhou, Sakamoto2020}. Most best performing models incorporated the information of the specific device upon training and evaluation. An exception was \cite{Lopez2020}, which treated the data for all devices jointly. In general, a broad variety of models and training schemes achieved scores over 90\%.\\
The 2021 \ac{UAD-S} edition also combined two datasets (MIMII DUE\cite{mimii_due} and ToyAdmos2\cite{toyadmos2}), extended in several aspects. Particularly, for each device, the 2021 dataset includes 7 {\it devices}, 6 {\it sections} and 2 {\it domains}, totalling 84 splits. Each device has 6 sections, which are balanced partitions of the data for evaluation purposes. Each section presents 2 domains: {\it source} and {\it target}, which differ in aspects like operating speed, machine load and environmental noise. As in 2020, the training data does not contain any anomalous sounds. The training data is also highly imbalanced: in all sections, only $\sim$0.3\% of the training samples are on the {\it target} domain. Test data is balanced in terms of devices, splits and domains. The evaluation procedure is similar to 2020, but this time an overall {\it score} is given as the harmonic mean across all \ac{AUC} and \ac{pAUC} scores\cite{2021t2_description}.\\
Out of 27 submissions for 2021, the \ac{AE} baseline ranked 21\textsuperscript{st} with a {\it score} of $\sim$56.4\% on the evaluation set. The 2021 winners\cite{LopezIL2021} ($\sim$66.8\%) propose a particularly heterogeneous ensemble, combining different ``{\it complementary}'' representations, objectives and models, rather than ``{\it relying on well-known domain adaptation techniques}''. A related concept is the contrastive loss applied by \cite{CaiSMALLRICE2021}, (9\textsuperscript{th} place, $\sim$61\%).  Second place was achieved by \cite{MoritaSECOM2021} ($\sim$65\%) with a simpler setup based on applying non-parametric inference methods (\ac{LOF} and \ac{KNN}) to trained embeddings. We note that, while it was observed that \ac{RepL}-based methods generally underperform reconstruction-based ones for \ac{UAD}\cite{shengchen_recons}, this does not seem to be the tendency here: reconstruction objectives are barely present in the top ranks, possibly due to sensitivity to domain shifts, and the emphasis is on representations, e.g.\ the importance of spectrogram hyperparameters noted by \cite{LopezIL2021} and the implications and effectiveness of different embeddings analyzed by \cite{WilkinghoffFKIE2021} (3\textsuperscript{rd} place, $\sim$64.2\%) which propose to use AdaCos\cite{adacos}. An exception is \cite{KuroyanagiNU-HDL2021} (4\textsuperscript{th} place, $\sim$63.75\%), which did propose a reconstruction-based method that compensates domain shift conditions.\\
Like in 2020, a variety of \ac{DL}-related approaches were adopted, but the scores were substantially lower in the 2021 edition. Keeping in mind the small differences in the evaluation procedure, we argue that the emphasis on \ac{RepL}-based methods, the relative success of non-parametric inference and the difficulty directly addressing domain shifts via well-known techniques point at the complexity of the task and the relevance of an adequate data representation, independently of the choice of model and training scheme. Therefore, in this exploratory work we propose to inspect the \ac{UAD-S} data distribution itself. The goal is to gain further insights in order to facilitate diagnosis and developement of further approaches.

\section{Inspecting representations with UMAP}
\label{sec:methodology}

Generally, direct exploration of high-dimensional data like that encountered in the discussed approaches is difficult. Fortunately, when data is organized in lower-dimensional structures it can be possible to retain some of its structure while projecting the data onto as few as 2 dimensions, allowing for informative visual inspection. \ac{UMAP}\cite{umap} is a non-linear projection technique that has been shown to surpass alternatives like \ac{PCA} and t-SNE\cite{tsne} in terms of speed, stability against reparametrizations and ``{\it meaningfulness}'' when applied on biological data\cite{umap_nature}. Nevertheless, dimensionality reduction usually entails information loss, and artifacts arise: dense clusters may appear spread out and well-separated structures may collide when projected. For this reason, we restrict ourselves to the {\bf assumption} that {\it if two regions appear separable on the \ac{UMAP} projection, they are also separable on the original representation}. Crucially, {\it the opposite is not necessarily true}: two regions appearing mixed could be due to a projection artifact. With this in mind, we focus on two \ac{UAD-S} properties:

\begin{itemize}
\item {\bf \acf{SEP}}: In a projection with ``good'' \ac{SEP}, a simple boundary can be drawn between anomalous and normal data with small error.
\item {\bf \acf{DSUP}}: If the training data provides set support for all normal data, and is separable from anomalous data, that set support can be directly used to discriminate anomalies. We consider that to be ``good'' \ac{DSUP}.
\end{itemize}
\begin{figure}
\includegraphics[width=1\columnwidth]{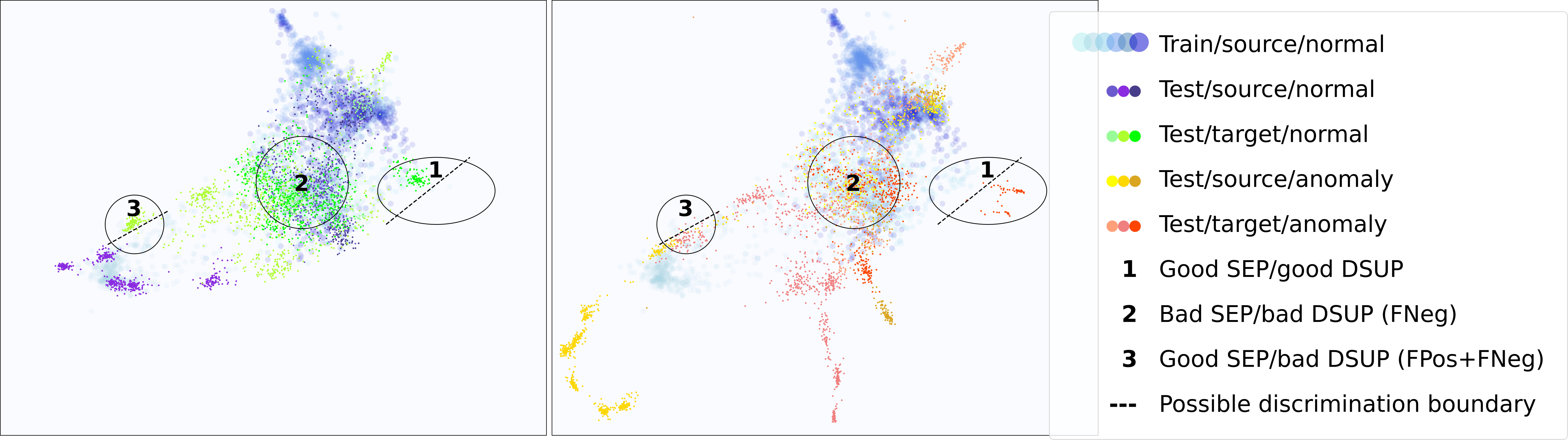}
\caption{Excerpt from the device \ac{UMAP} plot for {\it pump} with annotations in black. For example, region {\bf 1} presents good \ac{SEP} and \ac{DSUP}, since there is a simple boundary that clearly separates anomalies from normals and training data, and train/normal supports test/normal. {\it FPos} stands for false positives, and {\it FNeg} for false negatives. Each dot corresponds to 5 stacked log-mel frames. Color shades correspond to dataset {\it sections}. Training data is shown on both sides. Zoom to $\sim$1000\% for detail.}
\vspace{-10pt}
\label{fig:pump_dsup}
\end{figure}

Thus, by our assumption, if a given \ac{UMAP} projection presents good \ac{SEP} and \ac{DSUP}, we infer that the corresponding high-dimensional representation has a simple boundary to separate normal from anomalous data (i.e. good \ac{SEP}), and can be expressed using proximity to the training data (good \ac{DSUP}). We argue that this is beneficial for the \ac{UAD-S} task. Figure \ref{fig:pump_dsup} illustrates this.\\
The concept of \ac{SEP} could be quantified at a local level via e.g.\ \acp{SVM}\cite{svm}: given a set of data vectors $(\pmb{x}_1, \dots, \pmb{x}_N)$, labeled with -1 or 1 as $(y_1, \dots, y_N)$, the \ac{SVM} objective is to find the hyperplane parametrized by $\beta$ that creates the biggest margin between both classes, allowing for some error $\varepsilon$, via the following objective\cite[ch.\ 12]{stats}:
\begin{equation}
  \begin{aligned}
    \text{min} \norm{\beta}  \quad \text{s.t.} \, \left\{
    \begin{array}{ll}
      y_i(x_i^T \beta + \beta_o) \geq 1 - \xi_i\\
      \xi_i \geq 0\\
      \sum_i \xi_i \leq \varepsilon
    \end{array}
    \quad \forall i \in \{1, \dots, N\}
    \right.
  \end{aligned}
\end{equation}
Good \ac{SEP} would be then achieved when a simple boundary separates normal and anomalous data with low $\varepsilon$. \ac{DSUP} could be similarly quantified by comparing training and anomalous data, provided training data supports all normal data. But here we propose a complementary approach: to {\bf qualitatively} assess \ac{SEP} and \ac{DSUP} via visual inspection of \acp{UMAP}. To that end, we render a series of dual plots, showing {\bf anomalous test data on the right and normal test data on the left}.

\section{UMAP for DCASE 2021}
\label{sec:pipeline}

For our data sources, we merged the {\it Development} and {\it Additional Training} datasets\cite{2021t2_description} from \ac{DCASE} 2021, task 2. To illustrate the role of external datasets, we also incorporated the 10-second cut variant of Fraunhofer's IDMT-ISA-EE dataset\cite{isa}, and a custom subset of AudioSet consisting of 10-second segments from $\sim$40k unique videos. All audio files were converted to mono 16kHz, and $(-1, 1)$ normalization was applied. Based on high-performing systems from 2020, we computed amplitude spectrograms via the square modulus of \acp{STFT} with 1024 samples per window and 50\% overlap for all datasets. We then converted amplitudes to dB, yielding the log-STFTs. From the \ac{STFT} spectrograms, we also computed 128-bin melgrams\cite{mel} and converted them to dB, yielding the log-mels. We ended up with $\sim$300k frames per {\it source} split and 927 frames per {\it target} split, totaling $\sim$13 million for {\it source} and 39k for {\it target}. Our AudioSet subset had then $\sim$12.2 million frames, and Fraunhofer $\sim$223k. We used \texttt{librosa}\cite{librosa} for the above audio computations. We also computed 512-dimensional \ac{L3} embeddings with a hop size of 0.1 seconds using \texttt{openl3}\cite{l3_audio}. \ac{L3} embeddings encode longer-term relationships and this results in less frames (e.g.\ $\sim$3.5 million for our AudioSet). \acp{STFT} from environmental AudioSet videos were used for \ac{L3} audio training.\\
To encode temporal relationships, we stacked consecutive frames. In this work we explored 3 stack sizes: $1$, $5$ and $10$. We computed a set of 2D \ac{UMAP} projections for each of the 3 computed representations and 3 stack sizes. To get resolution at different scales, we computed one \ac{UMAP} per device plus a global \ac{UMAP}, totalling 72 \acp{UMAP}. Due to hardware limitations, the full datasets couldn't be processed and random samples were taken: For the per-device projections with stack size 1 and 5, we took 20k random samples for each validation split, and a maximum of 10k for every other split (recall that {\it target} training splits have just 927 frames). For the global projections with stack size 1 and 5, we took 2k samples per validation split, a maximum of 1k per training split, 50k for AudioSet and 50k for Fraunhofer. We also computed the stack size 10 \acp{UMAP} with no external data. For any given representation and stack size, we developed 3 kinds of scatter plots to enable different levels of detail: Global plots like Figure \ref{fig:global_stft} are based on the global \acp{UMAP} and show the full dataset, coloring the different devices. Device plots like Figure \ref{fig:toycar_l3} are based on the per-device \acp{UMAP} and color the different sections and domains. Section plots like Figure \ref{fig:valve_horn} are based on the per-device \acp{UMAP}, but they show a single section and color the specific audio files.

\section{Discussion}
\label{sec:discussion}

\begin{figure}
\includegraphics[width=1\columnwidth]{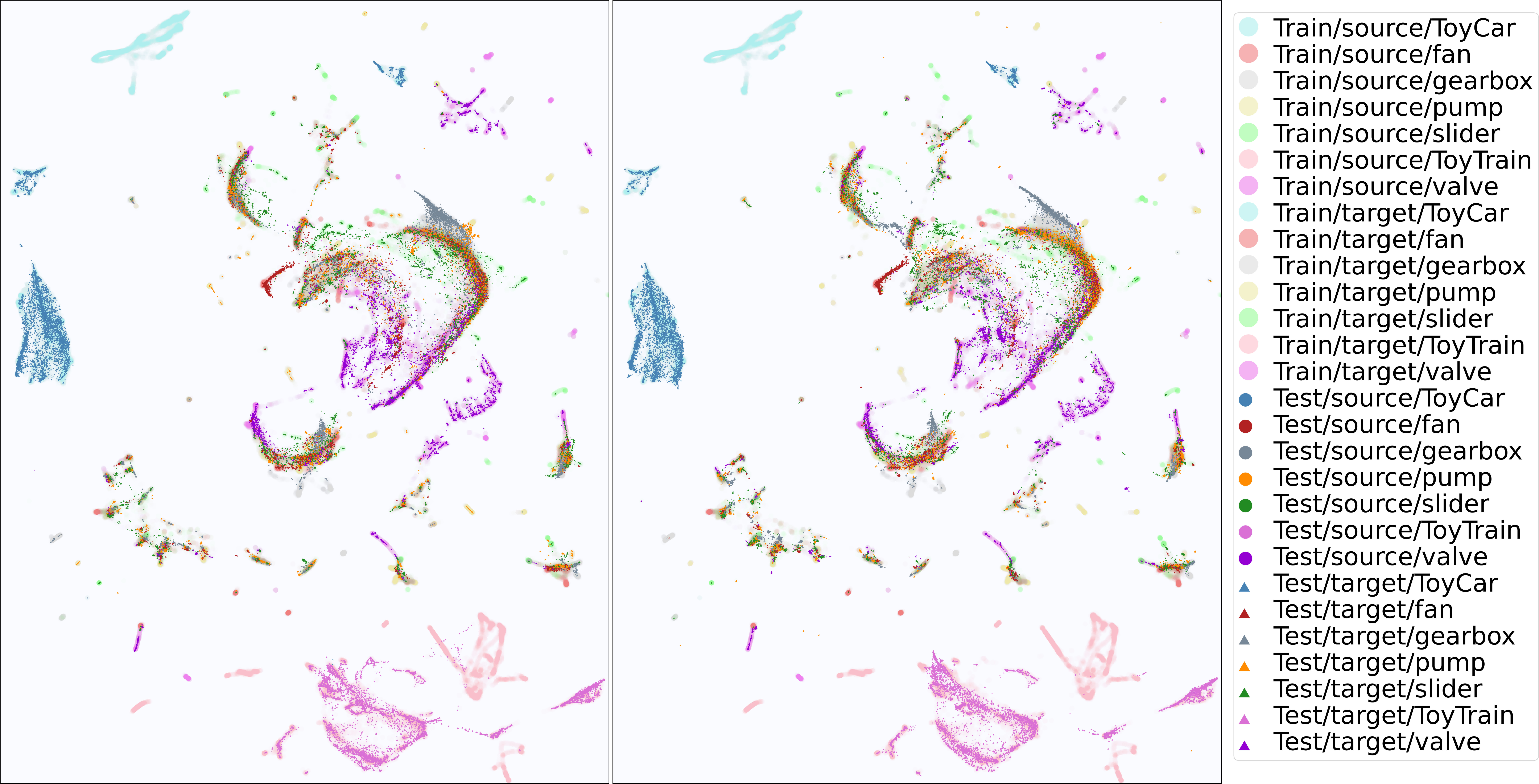}
\caption{Global \ac{UMAP} plot, sampled from the \ac{DCASE} 2021 dataset. Each dot corresponds to 10 stacked log-STFT frames. Smaller dots correspond to anomalies on the right, and normal data on the left. Training data is shown on both sides. Zoom to $\sim$1000\% for detail.}
\vspace{-5pt}
\label{fig:global_stft}
\end{figure}

In general, \ac{SEP} patterns across different devices and sections could be observed, but regions with both good \ac{SEP} and \ac{DSUP} were very hard to find, the best example we found has been already presented in Figure \ref{fig:pump_dsup}. Distinctively anomalous patterns are also scarce and do not appear to follow any obvious repeating patterns. We also observe that data from the {\it target} domain generally overlaps with the {\it source} domain.\\
The difference between ToyAdmos2 and MIMII DUE datasets can be seen at multiple scales: the ToyCar and ToyTrain clusters are clearly distinguishable from all other devices (see Figure \ref{fig:global_stft}), and the internal structure for the ToyAdmos2 devices is also apparently simpler than for MIMII DUE devices (compare Figures \ref{fig:toycar_l3} and \ref{fig:fan_l3}).\\
Another distinctive feature is the ``horn'' shape formed by the AudioSet data samples (e.g.\ Figures \ref{fig:valve_horn} and \ref{fig:fan_l3}). Since energy spectrograms are non-negative, they are confined to the first quadrant, which is a cone with a vertex on the zero-energy point. By checking the energies, we have observed that the lowest-energy samples are highly concentrated on all observed ``horns'' (a few outliers get projected elsewhere). This indicates that, accounting for the logarithmic conversion to dB, the representations conserve the conic geometry and the observed ``horn'' shape likely corresponds to the tip of the cone, giving a sense of origin that can aid interpretation.\\
Interestingly, in all \ac{L3} device plots for {\it fan}, the training data appears almost completely separated from the test data (see e.g.\ Figure \ref{fig:fan_l3}: the ``shadows'' have almost no test data on them). This means that a single \ac{L3} frame is enough to distinguish the {\it fan} training data from the test data fairly well. This should not be confused with a different phenomenon: some ``shadow'' clusters lack any overlying test data (e.g.\ the dark blue ones in Figure \ref{fig:toycar_l3}), but that is likely because those regions correspond to evaluation splits for which the challenge organizers did not release the test data. This is likely the case if the behavior is consistent across all representations.\\
Another particularity is that the AudioSet cluster appears to be smaller on the \ac{L3} representations, and the non-AudioSet data appears more scattered. This may be due to the fact that the \ac{L3} embeddings were trained on AudioSet and achieve a more compact representation there.\\
In the following we highlight several modelling {\bf hypotheses} based on the above observations and the literature. We refer to our online resources for extensive results and code.

\begin{enumerate}
\item {\bf Mixing ToyAdmos2 and MIMII DUE data may hinder performance}: Trivially distinguishable categories may lead to inefficient boundaries for anomaly discrimination. This was already proposed in \cite{primus2020}.

\item {\bf Temporal context and pretraining regulate a tradeoff between \ac{SEP} and \ac{DSUP}}: Generally, we observe that longer stack sizes provide better \ac{SEP}. This makes sense because given enough length all audio files can be uniquely identified. But we also observed that this tends to scatter data apart and worsen \ac{DSUP}. With pretrained embeddings, the observed tendency of concentrating the pretrained domain and scattering the rest may also entail a similar tradeoff. An ensemble with different tradeoff configurations may be beneficial. The complementary and contrastive approaches discussed in Section \ref{sec:dcase} may implicitly leverage this fact.

\item {\bf Normalization is a dominating factor}: If we interpret the data in Figure \ref{fig:valve_horn} as a cone with the vertex at the tip of the AudioSet ``horn'', renormalizing a frame would shift that frame roughly along the cone axis, which can greatly impact \ac{SEP} and \ac{DSUP}. The importance of proper normalization is supported by top-performing approaches like \cite{Lopez2020} and \cite{WilkinghoffFKIE2021}.

\item{\bf Incorporating domain-related priors may help performance}: Bad \ac{DSUP} only means that the the training data support can't be directly used for discrimination, but other kinds of prior knowledge still could be used to leverage the existing \ac{SEP}. The 2021 dataset provides {\it domain}-related labels describing the domain shifts in the training data that could be used as priors. To the best of our knowledge, none of the participants made use of it, and could be a beneficial addition.

\end{enumerate}
Lastly, we are aware of several methodological shortcomings:

\begin{enumerate}
\item We are only observing a subsample of the data, so extreme outliers are likely to be missed. Taking them into account may be crucial for successful analysis and detection.
\item As discussed in Section \ref{sec:methodology}, data projections can only be used to confirm \ac{SEP} and \ac{DSUP}, not to discard them.
\item Qualitative, visual inspection may also be subject to perceptual biases, e.g.\ by color strength or shape consistency. Furthermore, plotting anomalous and normal data on different sides hinders the visual detection of slight differences.
\item Encoding temporal relations by stacking successive frames can lead to suboptimal representations due to e.g.\ conditional relations among frames, normalization and weighting issues.
\end{enumerate}
Points 1 and 2 can be tackled by applying quantitative methods to the non-projected data, since the artifacts and size restrictions are imposed by the \ac{UMAP} step. To overcome any issues related to high data volume and dimensionality, \ac{LOF} and/or \ac{KNN}-based methods like the ones used in \cite{Durkota2020, arcface_zhou, MoritaSECOM2021} can be explored. Interactive exploration of the plots can help identifying small differences and overcoming perceptual biases. Representations that encode broader temporal context and other kinds of context can be explored to replace the frame stacks. Particularly, giving more weight to anomalous frames (or even ignoring very common frames) may help improving \ac{SEP} and \ac{DSUP}.

\begin{figure}
\includegraphics[width=1\columnwidth]{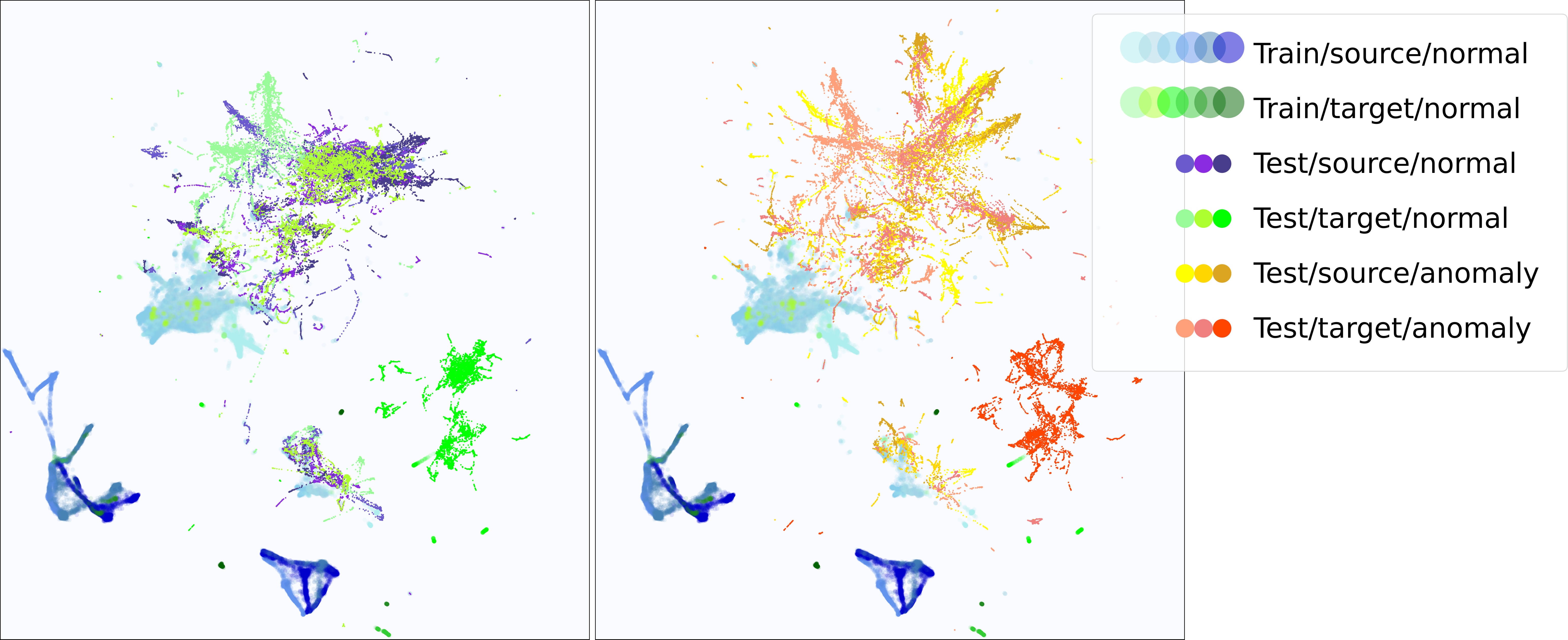}
\caption{Device \ac{UMAP} plot for {\it ToyCar}. Each dot corresponds to a single L3 frame. Color shades correspond to dataset {\it sections}. Training data is shown on both sides. Zoom to $\sim$1000\% for detail.}
\label{fig:toycar_l3}
\end{figure}

\begin{figure}
\includegraphics[width=1\columnwidth]{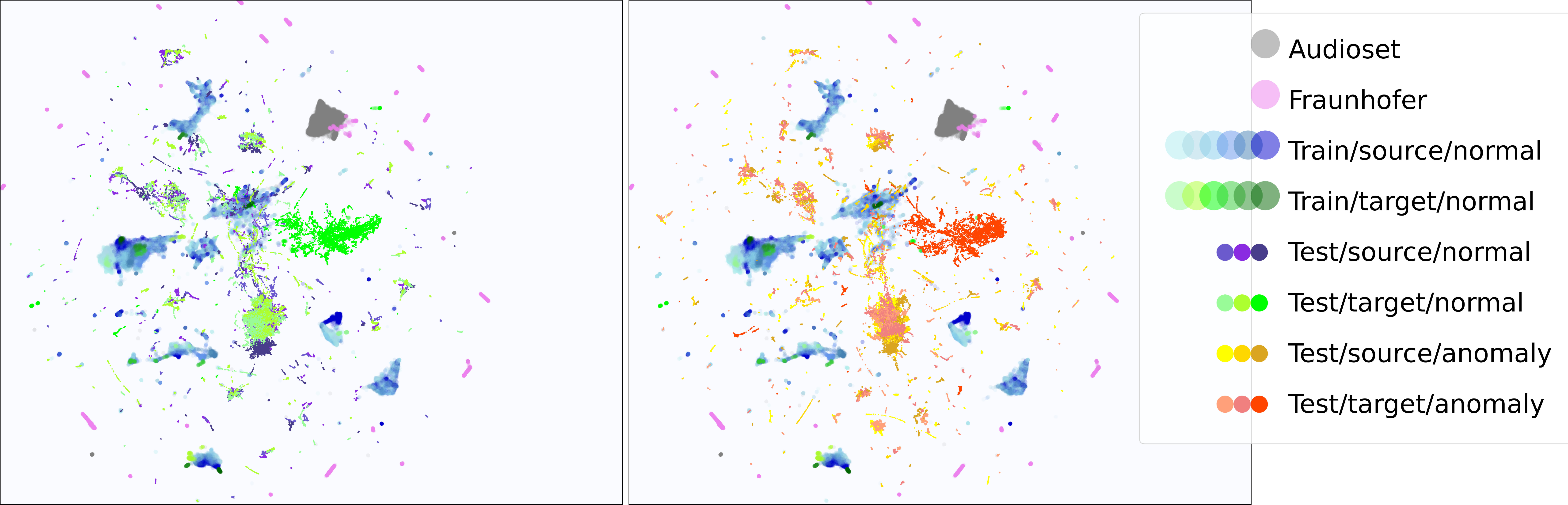}
\caption{Device \ac{UMAP} plot for {\it fan}. Each dot corresponds to a single L3 frame. Color shades correspond to dataset {\it sections}.  Training data is shown on both sides. Zoom to $\sim$1000\% for detail.}
\label{fig:fan_l3}
\end{figure}

\begin{figure}
  \includegraphics[width=1\columnwidth]{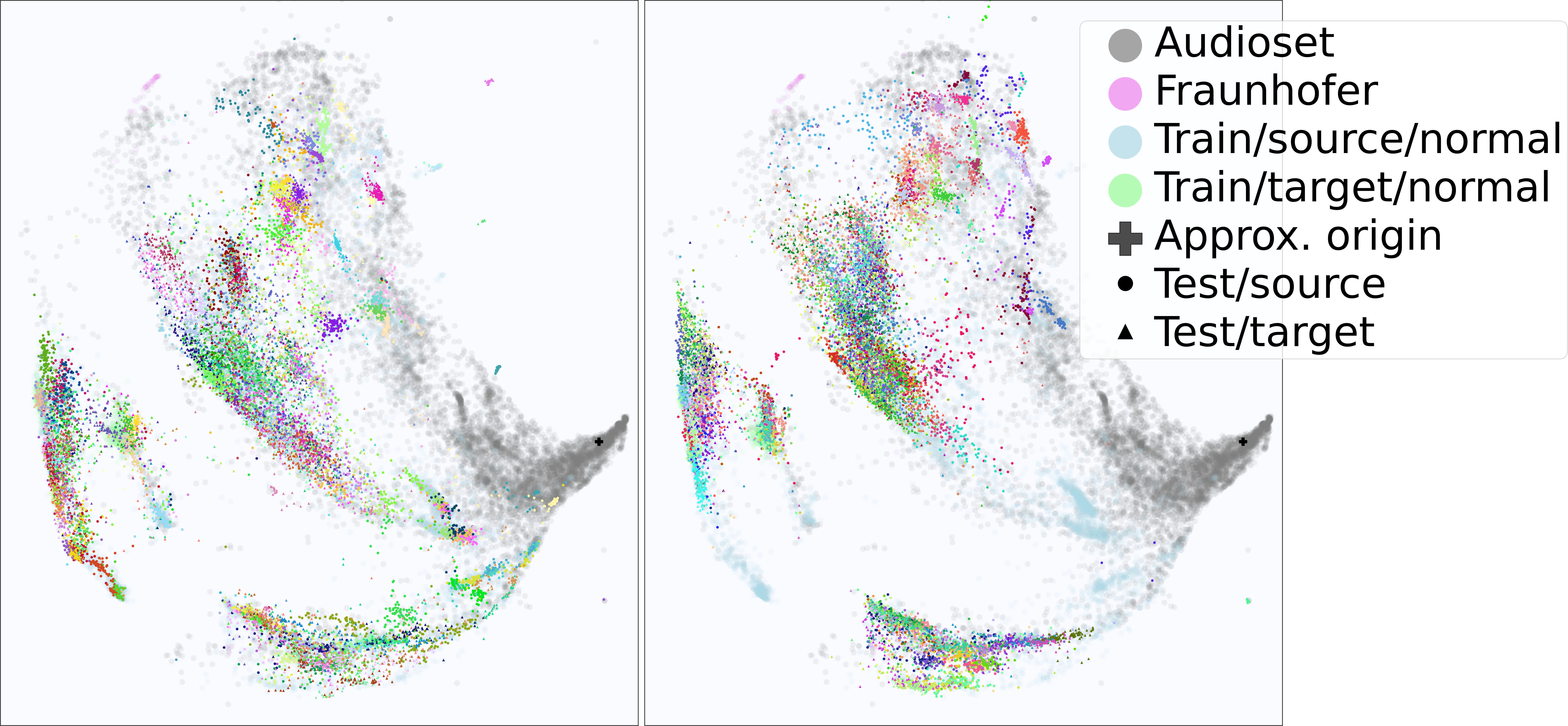}
  \caption{\ac{UMAP} plot for {\it valve}, section 0. Each dot corresponds to a single log-mel frame. Smaller dots correspond anomalies on the right, and normal data on the left, and are colored by audio filename. After ignoring the 500 lowest-energy frames, the cross signals the average position of the following 100 ones. Zoom to $\sim$1000\% for detail.}
  \label{fig:valve_horn}
\end{figure}

\section{Conclusion and Future Work}
\label{sec:conclusion}

In this paper we performed an analysis of fixed and learned \ac{UAD-S} data representations, based on the visual inspection of \acp{UMAP} and assessment of {\it separability} (\ac{SEP}) and {\it discriminative support} (\ac{DSUP}). In line with the difficulties encountered at the \ac{DCASE} challenge, we did not find consistently good \ac{SEP} and \ac{DSUP} in any of the observed representations. The representations helped to expose potential issues in connection with the literature, and ways to address them. Despite the discussed methodological shortcomings, we defend that visual \ac{UMAP} inspection can complement well other quantitative forms of analysis, and we hope that the software we provide can become a useful tool in the context of \ac{UAD-S}.\\
The analysis could be enhanced with interactive plots providing sonification (to better understand the topology by hearing it), and highlighting corresponding datapoints across different representations. Analysis of further representations and techniques like X-vectors and the Teager-Kaiser energy operator\cite{LopezIL2021} may also be of interest, as well as the impact of different embedding objectives on \ac{SEP} and \ac{DSUP}. Another possible extension could be to visualize the actual predictions of a system, extending this analyisis framework to supervised scenarios.\\

\section{Acknowledgments}

The authors would like to thank Helen Cooper for the support throughout the research process. This work was supported by grant EP/T019751/1 from the \ac{EPSRC} and made use of time on Tier 2 HPC facility JADE2, funded by \ac{EPSRC} grant EP/T022205/1.

% -------------------------------------------------------------------------
% Either list references using the bibliography style file IEEEtran.bst
\clearpage
\bibliographystyle{IEEEtran}

\bibliography{refs}

\end{sloppy}
\end{document}